# A Prospective Analysis of Security Vulnerabilities within Link Traversal-Based Query Processing (Extended Version)

Ruben Taelman and Ruben Verborgh

*IDLab, Department of Electronics and Information Systems, Ghent University – imec,*
*{firstname.lastname}@ugent.be*

**Abstract.**
The societal and economical consequences surrounding Big Data-driven platforms have increased the call for decentralized solutions. However, retrieving and querying data in more decentralized environments requires fundamentally different approaches, whose properties are not yet well understood. Link-Traversal-based Query Processing (LTQP) is a technique for querying over decentralized data networks, in which a client-side query engine discovers data by traversing links between documents. Since decentralized environments are potentially unsafe due to their non-centrally controlled nature, there is a need for client-side LTQP query engines to be resistant against security threats aimed at the query engine's host machine or the query initiator's personal data. As such, we have performed an analysis of potential security vulnerabilities of LTQP. This article provides an overview of security threats in related domains, which are used as inspiration for the identification of 10 LTQP security threats. Each threat is explained, together with an example, and one or more avenues for mitigations are proposed. We conclude with several concrete recommendations for LTQP query engine developers and data publishers as a first step to mitigate some of these issues. With this work, we start filling the unknowns for enabling querying over decentralized environments. Aside from future work on security, wider research is needed to uncover missing building blocks for enabling true decentralization.

## 1. Introduction

Contrary to the Web's initial design as a *decentralized* ecosystem, the Web has grown to be a very centralized place, as large parts of the Web are currently made up of a few large Bid Data-driven centralized platforms [1]. This large-scale centralization has lead to a number of problems related to personal information abuse, and other economic and societal problems. In order to solve these problems, there are calls to go back to the original vision of a decentralized Web. The leading effort to achieve this decentralization is Solid [1]. Solid proposes a radical *decentralization* of data across *personal data vaults*, where everyone is in full control of its own personal data vault. This vault can contain any number of documents, where its owner can determine who or what can access what parts of this data. In contrast to the current state of the Web where data primarily resides in a small number of huge data sources, Solid leads to a a Web where data is spread over a huge number of data sources.

Our focus in this article is not on decentralizing data, but on finding data after it has been decentralized, which can be done via *query processing*. The issue of query processing over data has been primarily tackled from a Big Data standpoint so far. However, if decentralization efforts such as Solid will become a reality, we need to be prepared for the need to query over a huge number of data sources. For example, decentralized social networking applications will need to be able to query over networks of friends containing hundreds or thousands of data documents. As such, we need new query techniques that are specifically designed for such levels of decentralization. One of the most promising techniques that could achieve this is called Link-Traversal-based Query Processing (LTQP) [2, 3]. LTQP is able to query over a set of documents that are connected to each other via *links*. An LTQP query engine typically starts from one or more documents, and *traverses* links between them in a crawling-manner in order to resolve the given query.

Since LTQP is still a relative young area of research, in which there are still a number of open problems that need to be tackled, notably result completeness and query termination [2]. Aside from these known issues, we also state the importance of *security*. Security is a highly important and well-investigated topic in the context of Web applications [4, 5], but it has not yet been investigated in the context of LTQP. As such, **we investigate in this article security issues related to LTQP engines**, which may threaten the integrity of the user's data, machine, and user experience, but also lead to privacy issues if personal data is unintentionally leaked. Specifically, we focus on data-driven security issues that are inherent to LTQP due to the fact that it requires a query engine to follow links on the Web, which is an uncontrolled, unpredictable and potentially unsafe environment. Instead of analyzing a single security threat in-depth, we perform a broader high-level analysis of multiple security threats.

Since LTQP is still a relatively new area of research, its real-world applications are currently limited. As such, we can not learn from security issues that arose in existing systems. Instead of waiting for –potentially unsafe– widespread applications of LTQP, we draw inspiration from related domains that *are* already well-established. Specifically, we draw inspiration from the domains of crawling and Web browsers in Section 2, and draw links to what impact these known security issues will have on LTQP query engines. In Section 3, we introduce a guiding use case that will be used to illustrate different threats with. After that, we discuss our method of categorizing vulnerabilities in Section 4. Next, we list 10 data-driven security vulnerabilities related to LTQP in Section 5, which are derived from known vulnerabilities in similar domains. For each vulnerability, we provide examples, and sketch possible high-level mitigations. Finally, we discuss the future of LTQP security and conclude in Section 6.

## 2. Related Work

This section lists relevant related work in the topics of LTQP and security.

### 2.1. Link-Traversal-Based Query Processing

More than a decade ago, Link-Traversal-based Query Processing (LTQP) [3, 2] has been introduced as an alternative query paradigm for enabling query execution over document-oriented interfaces. These documents are usually Linked Data [6] serialized using any RDF [7] serialization. RDF is suitable to LTQP and decentralization because of its global semantics, which allows queries to be written independently of the schemas of specific documents. In order to execute these queries, LTQP processing occurs over live data, and discover links to other documents via the *follow-your-nose principle* during query execution. This is in contrast to the typical query execution over centralized database-oriented interfaces such as SPARQL endpoints [8], where data is assumed to be loaded into the endpoint beforehand, and no additional data is discovered during query execution.

Concretely, LTQP typically starts off with an input query and a set of seed documents. The query engine then dereferences all seed documents via an HTTP GET request, discovers links to other documents inside those documents, and recursively dereferences those discovered documents. Since document discovery can be a very long (or infinite) process, query execution happens during the discovery process based on all the RDF triples that are extracted from the discovered documents. This is typically done by implementing these processes in an iterative pipeline [9]. Furthermore, since this discovery approach can lead to a large number of discovered documents, different reachabili-ty criteria [10] have been introduced as a way to restrict what links are to be followed for a given query.

So far, most research into LTQP has happened in the areas of formalization [10, 11], performance improvements [12, 13, 14], and query syntax [15]. One work has indicated the importance of *trustworthiness* [16] during link traversal, as people may publish false or contradicting information, which would need to be avoided or filtered out during query execution. Another work mentioned the need for LTQP engines to adhere to `robots.txt` files [17] in order to not lead to unintentional denial of service attacks of data publishers. Given the focus of our work on data-driven security vulnerabilities related to LTQP engines, we only consider this issue of *trustworthiness* further in this work, and omit the security vulnerabilities from a data publisher's perspective.

## 2.2. Vulnerabilities Of RDF Query Processing

Research involving the security vulnerabilities of RDF query processing has been primarily focused on injection attacks within Web applications that internally send SPARQL queries to a SPARQL endpoint. So far, no research has been done on vulnerabilities specific to RDF federated querying or link traversal. As such, we list the relevant work on single-source SPARQL querying hereafter.

The most significant type of security vulnerability in Web applications in general is *Injection through User Input*, of which SQL injection attacks [4] are a primary example. Orduna et al. [5] investigate this type of attack in the context of SPARQL queries, and show that *parameterized queries* can help avoid this type of attacks. A parameterized query is a query *template* that can contain multiple *parameters*, which can be instantiated with different values. To avoid injection attacks, parameterized query libraries will perform the necessary validation and escaping on the inserted values. The authors implemented parameterized queries in the Jena framework [18] as a mitigation example.

SemGuard [19] is a system that aims to detect injection attacks in both SPARQL and SQL queries for query engines that support both. A motivation of this work is that the use of parameterized queries is not always desirable, as systems may already have been implemented without them, and updating them would be too expensive. This approach is based on the automatic analysis of the incoming query's parse tree. It will check if the parse tree only has a leaf node for the expected user input, compared to the original template query's parse tree. If it does not have a leaf node, this means that the user is attempting to execute queries that were not intended by the application developer.

Asdhar et al. [20] analyzed injection attacks to Web applications via the SPARQL query language [21] and the SPARQL update language [22]. Furthermore, they provide *SemWebGoat*, a deliberately insecure RDF-based Web application for educational purposes around security. All of the discussed attacks involve some form of injection, leading to retrieval or modification of unwanted data, or denial-of-service by for example injecting the `?s ?p ?o` pattern. Such `?s ?p ?o` patterns cause all data to be fetched, which for large datasets can require long execution times, which may lead to denials of service for following SPARQL queries, or even crash the server and lead to availability issues [23].

## 2.3. Linked Data Access Control

Kirrane et al. [24] surveyed the existing approaches for achieving access control in RDF, for both authentication and authorization. The authors mention that only a minority of those works apply specifically to the document-oriented nature of Linked Data. They do however mention that non-Linked-Data-specific approaches could potentially be applied to Linked Data in future work. Hereafter, we briefly discuss the relevant aspects of access control research that applies to Linked Data. To the best of our knowledge, no security vulnerabilities have yet been identified for any of these.

### 2.3.1. Authentication

Authentication involves verifying an agent's identity through certain credentials. A WebID *(https://www.w3.org/wiki/WebID)* (Web Identity and Discovery) is a URL through which agents can be identified on the Web. WebID-TLS [25] is a protocol that allows authentication of WebID agents via TLS certificates. However, due to the limited support of such certificates in Web browsers, its usage is hindered. WebID-OIDC [26] is a more recent protocol is based on the OpenID Connect [27] protocol for authenticating WebID agents. Due to its compatibility with modern Web browsers, WebID-OIDC is frequently used inside the Solid ecosystem.

*2.3.2. Authorization*

Authorization involves determining who can read or write what kind of data. Web Access Control [28] is an RDF-based access control system that works in a decentralized fashion. It enables declarative access control policies for documents to be assigned to users and groups. Due to its properties, it is being used as default access control mechanism in the Solid ecosystem. Sacco et al. [29] extend Web Access Control to not only declare document-level access, but also on resource, statement and graph level. Costabello et al. [30] introduce the Shi3ld framework that enables access control for Linked Data Platform [31]. Two variants of this framework exist; one based on a SPARQL query engine, and one more limited variant that works without SPARQL queries. Kirrane et al. [32] introduce a framework for enabling query-based access control via query rewriting of simple graph pattern queries. Further, Steyskal et al. [33] provide an approach that is based on the Open Digital Rights Language. Finally, Taelman et al. [34] introduce a framework to optimize federated querying over documents that require access control, by incorporating authorizations into privacy-preserving data summaries.

*2.4. Web Crawlers*

Web crawling [35] is a process that involves collecting information on the Web by following links between pages. Web crawlers are typically used for Web indexing to aid search engines. Focused crawling [36] is a special form of Web crawling that prioritizes certain Web pages, such as Web pages about a certain topic, or domains for a certain country. LTQP can therefore be considered as an area of focused crawling that where the priority lies in achieving query results.

Web crawlers are often used for discovering vulnerable Web sites, for example through *Google Dorking* [37], which involves using Google Search to find Web sites that are misconfigured or use vulnerable software. Furthermore, crawlers are often used to find private information on Web sites. Such issues are however not the focus of this work. Instead, we are interested in the security of the crawling process itself, for which little research has been done to the best of our knowledge.

One related work in this area involves abusing crawlers to initiate attacks on other Web sites [38]. This may cause performance degradation on the attacked Web site, or could even cause the crawling agent to be blocked by the server. These attacks involve convincing the crawler to follow a link to a third-party Web site that exploits a certain vulnerability, such as an SQL injection. Additionally, this work describes a type of attack that allows vulnerable Web sites to be used for improving the PageRank [39] of an attacker-owned Web site via forged backlinks.

Some other works focus on mitigation of so-called *crawler traps* [40, 41] or *spider traps*. These are sets of URLs that cause an infinite crawling process, which can either be intentional or accidental. Such crawler traps can have multiple causes:
- Links between dynamic pages that are based on URLs with query parameters;
- Infinite redirection loops via using the HTTP 3xx range;
- Links to search APIs;
- Infinitely paged resources, such as calendars;
- Incorrect relative URLs that continuously increase the URL length.

Crawler traps are mostly discovered through human intervention when many documents in a single domain are discovered. Recently, a new detection technique was introduced [42] that attempts to measure the *distance* between documents, and rejects links to documents that are too similar.

*2.5. Web Browsers*

Web browsers enable users to visualize and interact with Web pages. This interaction is closely related to LTQP, with the main difference that LTQP works autonomously, while Web browsers are user-driven. Considering this close resemblance between these two domains, we give an overview of the main security vulnerabilities in Web browsers.

*2.5.1. Modern Web Browser Architecture*

Silic et al. [43] analyzed the architectures of modern Web browsers, determined the main vulnerabilities, and discuss how these issues are coped with.

Architecture-wise, browsers can be categorized into monolithic and modular browser architectures. The difference between the two is that the former does not provide isolation between concurrently executed Web programs, while the latter does. The authors argue that a modular architecture is important for security, fault-tolerance and memory management. They focused on the security aspects of the Chrome browser architecture [44], which consists of separate modules for the rendering engine, browser kernel, and plugins. Each of these modules is isolated in its own operating system process.

Silic et al. list the following main threats for Web browsers:

1. **System compromise**: Malicious arbitrary code execution with full privileges on behalf of the user. For example, exploits in the browser or third-party plugins caused by bugs. These types of attacks are mitigated through automatic updates once exploits become known.

2. **Data theft**: Ability to steal local network or system data. For example, a Web page includes a subresource to URLs using the file scheme (`file://`). which are usually blocked.

3. **Cross domain compromise**: Code from a Fully Qualified Domain Name (FQDN) executes code (or reads data) from another FQDN. For example, a malicious domain could extract authentication cookies from your bank's website you are logged into. This is usually blocked through the same-origin policy, but can be explicitly allowed through Cross-Origin Resource Sharing (CORS) *(https://fetch.spec.whatwg.org/#http-cors-protocol)*.

4. **Session hijacking**: Session tokens are compromised through theft or session token prediction. For example, cross-domain request forgery (CSRF) [45] is a type of attack that involves an attacker forcing a user logged in on another Web site to perform an action without their consent. Web browsers do not protect against these, but are typically handled by Web frameworks via the Synchronizer Token Pattern [46].

5. **User interface compromise**: Manipulating the user interface to trick the user into performing an action without their knowledge. For example, placing an invisible button in front of another button. This category also includes CPU and memory hogging to block the user from taking any further actions. Web browser have limited protections for these types of attacks that involve placing limitations on user interface manipulations.

*2.5.2. Lessons From Google Chrome*

Reis et al. [47] discuss on the three problems Google Chrome developers focus on to mitigate attacks:

1. **Reducing vulnerability severity**: In the real world, large projects such as Web browsers always contain bugs. Given this reality, Google Chrome consists of several sandbox layers reducing the damage should an exploit be discovered in one of the layers. The difficulty here lies in the fact that Web compatibility should be maintained, so that security restrictions do not break people's favorite Web sites.

2. **Reducing window of vulnerability**: If an exploit has been discovered, it should be patched as soon as possible. Google Chrome follows automated testing to ship security patches as soon as possible. All existing Chrome installations check for updates every five hours, and update in the background without disrupting the user experience.

3. **Reducing frequency of exposure**: In order to avoid people from visiting malicious Web sites for which the browser has not been patched yet, Google Chrome makes use of a continuously updating database of such Web sites. This will show a warning to the user before visiting such a site.

*2.5.3. Techniques For Mitigating Browser Vulnerabilities*

Browser Hardening [48] is based on the concept of reducing privileges of browsers to increase security. For example, browsers can be configured to disabled JavaScript and Adobe Flash, or whitelisted to trusted Web sites.

Fuzzing [49] is a technique that involves generating random data as input to software. Major Web browsers such as Google Chrome and Microsoft Edge perform extensive fuzzed testing by generating random Web pages and running them through the browser to detect crashes and other vulnerabilities.

*2.6. SQL Injection*

SQL injection attacks [4] are one of the most common vulnerabilities on Web sites where (direct or indirect) user input is not properly handled, and may lead to the attacker performing unintended SQL statements on databases. These types of attacks are typically mitigated through strong input validation, which are typically available in reusable libraries.

## 3. Use Case

In this section, we introduce a use case that will be used to illustrate the security threats discussed throughout this article.

We assume a Web with public and private information, which may for instance be achieved via personal data vaults following the principles of the Solid ecosystem [1]. This data vault is in full control of the owner, and they can host any kind of file in here, such as Linked Data files.

For this use case, we assume the existence of three people (Alice, Bob, and Carol), each having their own personal data vault. Alice uses her vault to store an address book containing the people she knows. Instead of storing contact details directly in the address book, she stores *links* to the profiles of her contacts (Bob and Carol). Bob and Carol can then self-define their own contact details. Fig. 1 shows an illustration of this setup.

The LTQP paradigm is well-suited to handle query execution over such setups. If Alice for instance would like to obtain the names of all her contacts, she could initiate a query starting from her address book as seed document, and the query engine would follow the links to her contacts, and obtain the names from their respective profiles. Some documents may require authentication before they can be accessed, for which Alice's query engine makes use of Alice's identity. In all threats throughout this article, we assume that Carol has malicious intentions that Alice is unaware of.

In this use case, two main roles can be identified. The first is the role of data publisher, which is taken up by Alice, Bob, and Carol though their person data vaults. The second is the role of the query initiator, which here applies to Alice, as she issues a query over her contacts.

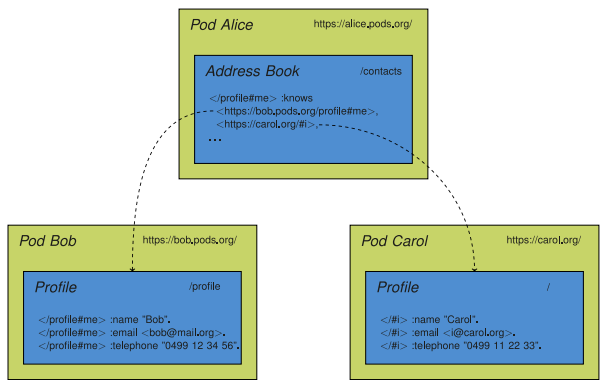

Fig. 1: Overview of the address book use case in which Alice has an address book with links to the profiles of Carol and Bob, which contain further details.

## 4. Classification of Security Vulnerabilities

In this section, we first introduce the background on classifying security vulnerabilities in software. After that, we introduce a classification method specifically for the LTQP domain, to assess the validity of our work.

### 4.1. Background

Security vulnerabilities in software can be classified using many different methods [50, 51]. Generic classification methods often result in very large taxonomies, which are shown to result in practical problems [50] because of their size and complexity.

Seacord et al. [50] claim that classification methods must be based on engineering analysis of the problem domain, instead of being too generic. For this, they suggest the use of domain-specific attributes for classifying security vulnerabilities for each domain separately. Furthermore, they introduce the following terminology for security vulnerabilities, by building upon earlier formal definitions of vulnerabilities [51]:

**Security flaw** A defect in a software application or component that, when combined with the necessary conditions, can lead to a software vulnerability.

**Vulnerability** A set of conditions that allows violation of an explicit or implicit security policy.

**Exploit** A technique that takes advantage of a security vulnerability to violate an explicit or implicit security policy.

**Mitigation** Techniques to prevent or limit exploits against vulnerabilities.

For the remainder of this article, we will make use of this terminology, and we adopt a method hereafter for classifying software vulnerabilities specific to the LTQP domain as recommended by Seacord et al. [50].

*4.2. Classification Method*

Our classification method considers the listing of several security *vulnerabilities*. Each vulnerability has two properties, as shown in Table 1. The *Possible exploits* property refers to a number of *exploits* that may take advantage of this vulnerability, and the *Mitigations* property refers to a number of *mitigations* that may prevent or limit this vulnerability. The different properties of each exploit are shown in Table 2, and the properties for each mitigation are shown in Table 3.

| Attribute | Values |
|---|---|
| Possible exploits | Intercepting private data, crashing a system, … |
| Mitigations | Sandboxing, same-origin policy, … |

Table 1: Vulnerability properties specific to LTQP, with several possible values for each attribute.

| Attribute | Values |
|---|---|
| Attacker | Data publisher, … |
| Victim | LTQP engine, query initiator, data publisher, … |
| Impact | Incorrect query results, system crash, … |
| Difficulty | Easy, medium, hard |

Table 2: Exploit properties specific to LTQP, with several possible values for each attribute.

| Attribute | Values |
|---|---|
| Location | LTQP engine, query initiator, data publisher, … |
| Difficulty | Easy, medium, hard |

Table 3: Mitigation properties specific to LTQP, with several possible values for each attribute.

## 5. Data-driven Vulnerabilities

As shown before in Subsection 2.2, most research on identifying security vulnerabilities within RDF query processing focuses on the query itself as a means of attacking, mostly through injection techniques. Since LTQP engines also accepts queries as input, these existing techniques will therefore also apply to LTQP engines.

In this work, we acknowledge the importance of these vulnerabilities, but we instead place our attention onto a new class of vulnerabilities that are specific to LTQP engines as a consequence of the open and uncontrolled nature of data on the Web. Concretely, we consider two main classes of security vulnerabilities to LTQP engines:

1. **Query-driven**: vulnerabilities that are caused by modifying queries that are the input to certain query engines.
2. **Data-driven**: vulnerabilities that are caused by the presence, structuring, or method of publishing data on the Web.

To the best of our knowledge, all existing work on security vulnerabilities within RDF query processing has focused on query-driven vulnerabilities. Given its importance for LTQP engines, we purely focus on data-driven vulnerabilities for the remainder of this work.

We identify three main orthogonal axes for security vulnerabilities, based on their exploit's potential impact area:

1. **Query Results**: vulnerabilities that lead to exploits regarding query results.
2. **Data Integrity**: vulnerabilities that lead to exploits regarding one or more user's data.
3. **Query Process**: vulnerabilities that lead to exploits regarding the stability of the query engine's process.

Table 4 gives an overview of all vulnerabilities that we consider in this article, and to what vulnerability axes they apply.

| Threat | Query Results | Data Integrity | Query Process |
|---|---|---|---|
| Unauthorized Statements | ✓ | | |
| Intermediate Result and Query Leakage | ✓ | | |
| Session Hijacking | | ✓ | |
| Cross-site Data Injection | ✓ | ✓ | |
| Arbitrary Code Execution | | ✓ | ✓ |
| Link Traversal Trap | | | ✓ |
| System hogging | | | ✓ |
| Document Corruption | | | ✓ |
| Cross-query Execution Interaction | | ✓ | |
| Document Priority Modification. | ✓ | | |

Table 4: An overview of all vulnerabilities related to LTQP that are considered in this article. They are decomposed into the different vulnerability axes to which they apply.

Hereafter, we explain and classify each vulnerability using the classification method from Section 4. For each vulnerability, we provide at least one possible example of an *exploit* based on our use case, and sketch at least one possible *mitigation*.

Unless mentioned otherwise, we do not make any assumptions about specific forms or semantics of LTQP, which can influence which links are considered. The only general assumption we make is that we have an LTQP query engine that follows links in any way, and executes queries over the union of the discovered documents.

*5.1. Unauthoritative Statements*

A consequence of the open-world assumption [52] where anyone can say anything about anything, is that both valid and invalid (and possibly malicious) things can be said. When a query engine is traversing the Web, it is therefore possible that it can encounter information that impacts the query results in an undesired manner. This information could be *untrusted* [16, 53], *contradicting*, or *incorrect*. Without mitigations to this vulnerability, query results from an LTQP can therefore never be really trusted, which brings the practical broad use of LTQP into question.

**Exploit: producing untrusted query results by adding unauthoritative triples**

Given our use case, Carol could for instance decide to add one additional triple to her profile, such as: `<https://bob.pods.org/profile#me> :name "Dave"`. She would therefore indicate that Bob's name is "Dave". This is obviously false, but she is "allowed" to state this under the open world assumption. However, this means that if Alice would naively query for all her friend's names via LTQP, she would have two names for Bob appear in her results, namely "Bob" and "Dave", where this second result may be undesired.

**Attacker** Data publisher (Carol)

**Victim** Query results from the LTQP engine of Alice

**Impact** Untrusted query results

**Difficulty** Easy (adding triples to an RDF document)

**Mitigation 1: applying content policies**

One solution to this vulnerability has been proposed [16], whereby the concept of *Content Policies* are introduced. These policies can capture the notion of what one considers authoritative, which can vary between different people or agents. In our example, Alice could for example decide to only trust her contacts to make statements about themselves, and exclude all other information they express during query processing. Such a policy would enable Alice's query for contact names to not produce Carol's false name for Bob. This concept of Content Policies does however only exist in theory, so no concrete mitigation to this vulnerability exist yet.

**Location** Data publishers and LTQP engines

**Difficulty** Currently hard (content policy implementations do not exist yet)

**Mitigation 2: tracking provenance**

Another solution to this vulnerability has been suggested [53] to make use of data provenance [54]. In contrast to the previous mitigation, this approach would not limit what is incorporated from what sources, but instead it would document the sources information came from. The end-user can then decide afterwards what provenance trails it seems trustworthy.

**Location** LTQP engines

**Difficulty** Medium

*5.2. Intermediate Result And Query Leakage*

This vulnerability assumes the existence of a *hybrid* LTQP query engine that primarily traverses links, but can exploit database-oriented interfaces such as SPARQL endpoints if they are detected in favour of a range of documents. Furthermore, we assume a range of documents that require authentication, as their contents are not accessible to everyone. Query engines typically decompose queries into smaller sub-queries, and join these intermediate results together afterwards. In the case of a hybrid LTQP engine, intermediate results that are obtained from the traversal process from non-public documents could be joined with data from a discovered SPARQL endpoint. An attacker could therefore set up an interface that acts as a SPARQL endpoint, but is in fact a **malicious interface that intercepts intermediate results** from LTQP engines.

**Exploit: capturing intermediary results via malicious SPARQL endpoint**

Based on our use case, Carol could include a triple with a link to the SPARQL endpoint at `http://attacker.com/sparql`. If Alice makes use of a hybrid LTQP engine with an adaptive query planner, this internal query planner could decide to make use of this malicious endpoint once it has been discovered. Depending on the query planner, this could mean that non-public intermediate results from the traversal process such as Bob's telephone are used as input to the malicious SPARQL endpoint. Other query planning algorithms could even decide to send the full original SPARQL query into the malicious endpoint. Depending on the engine and its query plan, this could give the attacker knowledge of intermediate results, or even the full query. This vulnerability enables attackers to do obtain insights to user behaviour, which is a privacy concern. A more critical problem is when private data is being leaked that normally exists behind access control, such as bank account numbers.

**Attacker** SPARQL endpoint publisher (Carol)

**Victim** Intermediary results of the LTQP engine of Alice

**Impact** Leakage of (intermediary) query results

**Difficulty** Medium (setting up a malicious SPARQL endpoint)

**Mitigation: Same-origin policy**

As this vulnerability is similar to the *cross-domain compromise* and *data theft* vulnerabilities in Web browsers [43]. A possible solution to it would be in the form of the *same-origin policy* that is being employed in most of today's Web browsers. In essence, this would mean that intermediate results can not be used across different Fully Qualified Domain Names (FQDN). Such a solution would have to be carefully designed as to not lead to performance issues, or lead to significant querying restrictions that would lead to fewer relevant query results. A mechanism in the form of Cross-Origin Resource Sharing (CORS) *(https://fetch.spec.whatwg.org/#http-cors-protocol)* could be used as a workaround to explicitly allow intermediate result sharing from one a domain to another. Such a workaround should be designed carefully, as not to suffer from the same issues as CORS [55]. Related to this, just like Web browsers, query engines may provide queryable access to local files using the `file://` scheme. Web browsers typically block requests to these from remote locations due to their sensitive nature. Similarly, query engines may decide to also block requests to URLs using the `file://` scheme, unless explicitly enabled by the user. This approach is for example followed by the Comunica query engine [56].

**Location** LTQP engines

**Difficulty** Easy (hard with CORS-like workaround)

*5.3. Session Hijacking*

In this vulnerability, we assume the presence of some form of authentication (such as WebID-OIDC [26]) that leads to an active authenticated session. This vulnerability is similar to that of Web browsers, where the session token can be compromised through theft or session token prediction. Such a vulnerability could lead to cross-domain request forgery (CSRF) [45] attacks, where an attacker forces the user to perform an action while authenticated without the user's consent.

**Exploit: triggering unintended operations on SPARQL endpoint behind access control**

For example, we assume that Alice has a flawed SPARQL endpoint running at `http://my-endpoint.com/sparql`, which requires Alice's session for accepting read and write queries. Alice's query engine may have Alice's session stored by default for when she wants to query against her own endpoint. If Carol knows this, she could a malicious triple with a link to `http://my-endpoint.com/sparql?query=DELETE * WHERE { ?s ?p ?o }` in her profile. While the SPARQL protocol [8] only allows update queries via HTTP POST, Alice's flawed query engine could implement this incorrectly so that update queries are also accepted via HTTP GET. If Alice executes a query over her address book, the query engine could dereference this link with her session enabled, which would cause her endpoint to be cleared. This vulnerability is however not specific to SPARQL endpoints, but may occur on any type of Web API that allows modifying data via HTTP GET requests.

**Attacker** Data publisher (Carol)

**Victim** Alice's stored data

**Impact** Removal or modification of Alice's stored data

**Difficulty** Easy (adding a malicious link to flawed endpoint)

**Mitigation: same-origin policy**

This vulnerability should be tackled on different fronts, and primarily requires secure and well-tested software implementations. First, it is important that authentication-enabled query engines do not leak sessions across different origins. This could be achieved by scoping authentication sessions to the origin URL in which they were created, and for each document only allow sessions to be used that are contained within the scope of the document's origin.

**Location** LTQP engines

**Difficulty** Medium

**Mitigation: only handle HTTP GET during traversal**

A second mitigation is that traversal should only be allowed using the HTTP GET method. This may not always be straightforward, as hypermedia vocabularies such as Hydra [57] allow specifying the HTTP method that is to be used when accessing a Web API (e.g. `hydra:method`). Given an unsecure query engine implementation, such HTTP method declarations could be exploited.

**Location** LTQP engines

**Difficulty** Medium

**Mitigation: adhere to read-only semantics of HTTP GET**

A third mitigation is that Web APIs must strictly follow the read-only semantics of HTTP GET, which is not always followed by many Web APIs [58], either intentionally or due to software bugs.

**Location** Data publishers

**Difficulty** Medium

*5.4. Cross-Site Data Injection*

This vulnerability concerns ways by which attackers can inject data or links into documents. For instance, HTTP GET parameters are often used to parameterize the contents of documents. If such parameters are not properly validated or escaped, they can be used by attackers to include malicious data or links.

**Exploit: injecting untrusted links via flawed trusted API**

For example, assuming Alice executes a query over a page from Carol, and a compromised API `http://trusted.org/?name` that dynamically creates RDF responses based on the `?name` HTTP GET parameters. In this case, the API simply has a Turtle document template into which the name is filled in as a literal value, but it does not do any escaping. We assume Alice decides to fully trust all links from `http://trusted.org/` to other pages, but only trust information directly on Carol's page or links to other trusted domains. If Carol includes a link to `<http://trusted.org/?name=Bob". <> rdfs:seeAlso <http://hacker.com/invalid-data>. <> foaf:name "abc"`", then this would cause the API to produce a Turtle document that contains a link to `http://hacker.com/invalid-data`, which would lead to unwanted data to be included in the query results.

**Attacker** Data publisher (Carol)

**Victim** Query results from the LTQP engine of Alice

**Impact** Untrusted query results

**Difficulty** Easy (adding triples to an RDF document)

**Mitigation: validate API parameters**

No single technique can fully mitigate this vulnerability. Just like SQL injection attacks [4] on Web sites, Web APIs should take care of input validation, preferably via reusable and rigorously tested software libraries.

**Location** Data publishers

**Difficulty** Medium

**Mitigation: expressive content policies**

On the side of query engines, this vulnerability may partially mitigated by carefully designing content policies. In the case of our example, defining a policy that enables the full range of (direct and indirect) links to be followed from a single domain can be considered unsafe. Instead, more restrictive policies may be enforced, at the cost of expressivity and flexibility.

**Location** LTQP engines

**Difficulty** Currently hard (content policies do not exist yet)

*5.5. Arbitrary Code Execution*

Advanced crawlers such as the Googlebot [59] allow JavaScript logic to be executed for a limit duration, since certain HTML pages are built dynamically via JavaScript at the client-side. In this vulnerability, we assume a similar situation for LTQP, where Linked Data pages may also be created client-side via an expressive programming language such as JavaScript. This would in fact already be applicable to HTML pages that dynamically produce JSON-LD script tags or RDFa in HTML via JavaScript. In order to query over such dynamic Linked Data pages, a query engine must initiate a process similar to Googlebot's JavaScript execution phase. Such a process does however open the door to potentially major security vulnerabilities if malicious code is being read and executed by the query engine during traversal.

**Exploit: manipulate local files via overprivileged JavaScript execution**

For example, we assume that Alice's LTQP query engine executes JavaScript on HTML pages before extracting its RDFa and JSON-LD. Furthermore, this LTQP engine has a security flaw that allows executed JavaScript code to access and manipulate the local file system. Carol could include a malicious piece of JavaScript code in her profile that makes use of this flaw to upload all files on the local file system to the attacker, and deletes all files afterwards so that she can hold Alice's data for ransom.

**Attacker** Data publisher (Carol)

**Victim** Files on machine in which Alice's query engine runs

**Impact** Removal or modification of files on Alice's machine

**Difficulty** Easy (adding JavaScript code to a document)

**Mitigation: sandbox code execution**

One of the problems Google Chrome developers focus on is *reducing vulnerability severity*, which involves running logic inside one or more sandboxes to reduce the chance of software bugs to lead to access to more critical higher-level software APIs. While software bugs are nearly impossible to avoid in real-world software, a similar sandboxing approach helps reducing the severity of attacks involving arbitrary code execution. Such a sandbox would only allow certain operations to be performed, which would not include access to the local file system. If this sandbox would also support performing HTTP requests, then the *same-origin policy* should also be employed to mitigate the risk of cross-site scripting (XSS) attacks.

**Location** LTQP engines

**Difficulty** Medium

*5.6. Link Traversal Trap*

LTQP by nature depends on the ability of iteratively following links between documents. It is however possible that such **link structures cause infinite traversal paths** and make the traversal engine get trapped, either intentionally or unintentionally, just like crawler traps. Given this reality, LTQP query engines must be able to detect such traps. Otherwise, query engines could never terminate, and possibly even produce infinite results.

**Exploit: forming a link cycle**

A link cycle is a simple form of link traversal trap that could be formed in different ways. First, at application-level, Carol's profile could contain a link path to document X, and document X could contain a link path back to Carol's profile. Second, at HTTP protocol-level, Carol's server could return for her profile's

URL an (HTTP 3xx) redirect chain to URL X, and URL X could contain a redirect chain back to the URL of her profile. Third, at application level, a cycle structure could be simulated via virtual pages that always link back to similar pages, but with a different URL. For example, the Linked Open Numbers [60] project generates a long virtual sequence of natural numbers, which could produce a bottleneck when traversed by an LTQP query engine.

**Attacker** Data publisher (Carol)

**Victim** Query process of Alice's query engine

**Impact** Unresponsiveness of Alice's query engine

**Difficulty** Easy

**Mitigation: tracking history of links**

Problems with first and second form of link cycles could be mitigated by letting the query engine keep a history of all followed URLs, and not dereference a URL that has already been passed before. The third form of link cycle makes use of distinct URLs, so this first mitigation would not be effective.

**Location** LTQP engines

**Difficulty** Easy

**Mitigation: limit link path length**

An alternative approach that would mitigate this third form –and also the first two forms at a reduced level of efficiency–, is to place a limit on the link path length from a given seed document. For example, querying from page 0 in the Linked Open Number project with a link path limit of 100 would cause the query engine not to go past page 100. This is the approach that is employed by the recommended JSON-LD 1.1 processing algorithm [61] for handling recursive `@context` references in JSON-LD documents. HTTP libraries typically also limit the number of redirects at protocol-level, e.g. the `maxRedirects` option in the `follow-redirects` *(https://github.com/follow-redirects/follow-redirects)* library that is set to a default value of 21. Different link path limit values could be applicable for different use cases, so query engines could consider making this value configurable for the user.

**Location** LTQP engines

**Difficulty** Easy

**Mitigation: measuring document similarity**

Other more advanced mitigation techniques from the domain of crawler trap mitigation could be extended, such as the one that measures similarities between documents to detect crawler traps with common structures [42]. For crawler traps that do not share commonalities across documents, mitigation techniques do not exist yet to the best of our knowledge.

**Location** LTQP engines

**Difficulty** Hard

*5.7. System Hogging*

The *user interface compromise* vulnerability for Web browsers includes attacks involving CPU and memory hogging through (direct or indirect) malicious code execution or by exploiting software flaws. Such vulnerabilities also exist for LTQP query engines, especially regarding the use of different RDF serializations, and their particularities with respect to parsing.

**Exploit: producing infinite RDF documents**

For example, RDF serializations such as Turtle [62] are implicitly designed as to allow streaming serialization and deserialization. JSON-LD even explicitly allows this through its Streaming JSON-LD note [63]. Due to this streaming property, RDF documents of infinite size can be generated, since serializations place no limits on their document sizes. Valid use cases exist for publishers to generate infinite RDF documents, which can be streamed to query engines. Query engines with non-streaming or flawed streaming parsers, can lead to CPU and memory issues. Furthermore, similar issues can occur due to very long or infinite IRIs or literals inside documents. Other attacks could exist that specifically target known flaws in RDF parsers that cause CPU or memory issues.

**Attacker** Data publisher (Carol)

**Victim** Machine in which Alice's query engine runs

**Impact** Unresponsiveness or crashing of Alice's query engine or machine

**Difficulty** Easy

**Mitigation: placing limits for RDF syntaxes**

Even though typically omitted from RDF format specifications, implementations often place certain limits on maximum document, IRI and literal lengths. For instance, SAX parsers [64] typically put a limit of 1 megabyte on IRIs and literals, and provide the option to increase this limit when certain documents would exceed this threshold.

**Location** LTQP engines

**Difficulty** Medium (identifying all possible limits may not be trivial)

### Mitigation: sandbox RDF parsing

Applying the approach of sandboxing on RDF parsers would also help mitigate such attacks, by for example placing a time and memory limit on the parsing of a document. If LTQP engines would allow arbitrary code execution, then more extensive system hogging mitigations would be needed just like in Web browsers [43].

**Location** LTQP engines

**Difficulty** Medium

*5.8. Document Corruption*

Since the Web is not a centrally controlled system, it is possible that documents are incorrectly formatted, either intentional or unintentional. RDF formats typically prescribe a restrictive syntax, which require parsers to emit an error when it encounters illegal syntax. When an LTQP engine discovers and parses a large number of RDF documents, possibly in an uncontrolled manner, it is undesired that a syntax error in just a single RDF document can cause the whole query process to terminate with an error. Furthermore, the phenomenon of *Link Rot* [65] can lead to links going dead (HTTP 404) at any point in time, while finding a link to a URL that produces a 404 response should not always cause the query engine to terminate.

### Exploit: publishing an invalid RDF document

For example, Carol could decide to introduce a syntax error in her profile document, or she could simply remove it to produce a 404 response. This would could cause Alice's queries over her friends from that point on to fail.

**Attacker** Data publisher (Carol)

**Victim** Alice's query engine

**Impact** Crashing of Alice's query engine

**Difficulty** Easy

### Mitigation: sandbox RDF parsing

The sandbox approach is well-suited for handling these types of attacks. RDF parsing for each document can run in a sandbox, where errors in this document would simply cause parsing of this document to end without crashing the query engine. Optionally, a warning could be emitted to the user. The same approach could be followed for HTTP errors on the protocol level, such as HTTP 404's. This approach is followed by the Comunica query engine [56] via its lenient execution mode.

**Location** LTQP engines

**Difficulty** Medium

### Mitigation: lenient RDF parsing

An alternative mitigation would be to create more lenient RDF parsers that accept syntax errors and attempts to derive the intended meaning, similar as to how (non-XHTML) HTML parsers are created. The downside of this is that such parsers would not strictly adhere to their specifications.

**Location** LTQP engines

**Difficulty** Hard

*5.9. Cross-Query Execution Interaction*

Query engines of all forms typically make use of caching techniques to improve performance of query execution. LTQP query engines can leverage caching techniques for document retrieval. Within a single query execution, or across multiple query executions, the documents may be reused, which could reduce the overall number of HTTP requests. Such forms of caching can lead to vulnerabilities based on information leaking across different query executions. We therefore make the assumption of caching-enabled LTQP engines in this vulnerability.

### Exploit: timing attack to determine prior knowledge

A first exploit of this vulnerability is an attack that enables Carol to gain knowledge about whether or not Bob's profile has been requested before by Alice. We assume that the Alice's engine issues a query over a document from Carol listing all her pictures. We also assume that Bob's profile contains a link to Carol's profile. If Carol includes a link from her pictures document to Bob's profile, and Bob's profile already links to Carol's profile, then the query engine could fetch these three documents in sequence (Carol's pictures,

Bob's profile, Carol's profile). Since Carol's pictures and profile are in control of Carol, she could perform a timing attack [66] to derive how long the Alice's query engine took to process Bob's profile. Since HTTP delays typically form the bottleneck in LTQP, Carol could thereby derive if Bob's profile was fetched from a cache or not. This would enable Carol to gain knowledge about prior document lookups, which could for example lead to privacy issues with respect to the user's interests.

**Attacker** Data publisher (Carol)

**Victim** Privacy about Alice's document usage

**Impact** Alice's document usage becomes known to Carol

**Difficulty** Hard

### Exploit: unauthenticated cache reuse

A second exploit assumes the presence of a software flaw inside Alice's LTQP query engine that makes document caches ignore authorization information. This example is also a form of the *Intermediate Result and Query Leakage* vulnerability that was explained before, for which we assume the existence of a *hybrid* LTQP query engine. If Alice queries a private file containing her passwords from a server using its authentication key, this can cause this passwords file to be cached. If Carol has a query endpoint that is being queried by Alice, and Carol is aware of the location of Alice's passwords, then she could maliciously introduce a link to Alice's passwords file. Even if the query was not executed with Alice's authentication key, the bug in Alice's query engine would cause the passwords file to be fetched in full from the cache, which could cause parts of it to be leaked to Carol's query endpoint.

**Attacker** Data publisher (Carol)

**Victim** Alice's private data

**Impact** Alice's private data is leaked

**Difficulty** Easy (if cache is flawed)

### Mitigation: sandboxing query execution

In order to mitigate this vulnerability, the isolation model that is used in Web browsers [43] could be reused. When applied to LTQP query engines, this could mean that each query would be executed in a separate sandbox, so that information can not leak across different query executions. A downside of this approach is that this may cause a significant performance impact when similar queries are executed in sequence, and would cause identical documents to not be reused from the cache anymore. In order to mitigate this drawback, solutions may be possible to allow "related queries" to be executed inside the same sandbox.

**Location** LTQP engines

**Difficulty** Medium

*5.10. Document Priority Modification*

Different techniques are possible to determine the priority of documents [12] during query processing. If queries do not specify a custom ordering, this prioritization will impact the ordering of query results. Some of these techniques are purely graph-based, such as PageRank [39], and can therefore suffer from purely data-driven attacks. This vulnerability involves attacks that can influence the priority of documents, and thereby maliciously influence what query results come in earlier or later.

### Exploit: malicious PageRank prioritization of documents

One possible exploit is similar to the attack to modify priorities within crawlers [38]. We assume that Alice issues a query that returns grocery stores in the local area, which is executed via a LTQP query engine that makes use of PageRank to prioritize documents. Furthermore, we assume a highly-scoring, but vulnerable API that accepts HTTP GET parameters that can be abused to inject custom URLs inside the API responses. If Carol aims to increase the ranking of her grocery store within Alice's query for better visibility, then she could exploit this vulnerable API. Concretely, Carol could place links from the grocery store's page to this vulnerable API using GET parameters that would cause it to link back to Carol's grocery store. Such an attack would lead to a higher PageRank for Carol's grocery store, and therefore an earlier handling and result representation of Carol's grocery store.

**Attacker** Data publisher (Carol)

**Victim** Order of Alice's query results

**Impact** Carol's page is ranked higher

**Difficulty** Medium

### Mitigation: validate API parameters

Several mitigations have been proposed for these types of attacks [38]. A first solution is to place responsibility at the API, and expecting it to patch the exploit.

**Location** Data publishers

**Difficulty** Medium

### Mitigation: content policies

A second mitigation involves publishers to expose policies that explicitly authorize what links should be considered legitimate, and LTQP query engines inspecting these policies when determining document priorities.

**Location** Data publishers and LTQP engines

**Difficulty** Currently hard (content policies do not exist yet)

### Mitigation: automated learning of legitimate links

A third mitigation is to use machine-learning to distinguishing non-legitimate from legitimate links. A combination of the three approaches can be used to mitigate this vulnerability.

**Location** LTQP engines

**Difficulty** Hard

## 6. Conclusions

In this article, we have identified ten prospective security vulnerabilities related to LTQP, inspired by known vulnerabilities in related domains. For each vulnerability, we proposed one or more avenues for mitigations.

Some of these vulnerabilities can already be partially tackled through existing security vulnerability mitigation techniques aimed at both *LTQP engine developers* and *data publishers*. As such, we **recommend LTQP engine developers to**:
- apply the **same-origin policy** for authentication sessions (Subsection 5.3);
- only allow traversal using the **HTTP GET** method (Subsection 5.3);
- **restrict link path lengths** to avoid link traversal traps (Subsection 5.6);
- run untrusted code and RDF parsing over untrusted data in a **sandbox** (Subsection 5.5, Subsection 5.7);
- make errors in the sandbox **not crash the query process** (Subsection 5.8).

At the same time, **recommend data publishers to**:
- **validate input** to avoid data injection (Subsection 5.4);
- ensure **HTTP GET** requests are **read-only** (Subsection 5.3).

For the following security vulnerabilities, **no concrete mitigation techniques exist yet**:
- Unauthorized Statements (Subsection 5.1)
- Intermediate Result and Query Leakage (Subsection 5.2)
- Cross-query Execution Interaction (Subsection 5.9)
- Document Priority Modification Subsection 5.10

With this prospective analysis, we have illustrated the importance of more security-oriented research in the domain on LTQP and the general handling of decentralized environments such as Solid [1], especially in presence of data behind authentication. While some of these vulnerabilities can be mitigated using existing techniques in related domains, further research on them is needed to test their impact on implementation, analyze their performance impact, introduce more performant techniques and algorithms, and introduce and apply attack models to test their effectiveness. Furthermore, for the security vulnerabilities for which no concrete mitigations exist yet, research is perhaps even more critical. Since our analysis of security vulnerabilities is by no means exhaustive, additional research efforts are needed to uncover and predict potential security vulnerabilities in LTQP. Such future research—with our work as a first step—is crucial for enabling a decentralized Web which we can query securely.